\documentstyle[preprint,aps,floats,epsfig]{revtex} % For drafts and producing CLNS reports

\begin{document}

\draft % makes pacs numbers print

\tighten % To save trees, please use this for the copy to be distributed! 
\preprint{\vbox{\hbox{CLNS 97-1484 \hfill}
                \hbox{CLEO 97--10  \hfill}}}

\def\today{June 2, 1997}
 
\title{A New Upper Limit on the Decay $\eta \rightarrow e^{+}e^{-}$}
\date{\today}
\maketitle
\begin{abstract}
We have searched for the rare decay of the eta meson $\eta \rightarrow
e^{+}e^{-}$ using the CLEO II detector.  The $\eta$'s were produced in
$e^{+}e^{-}$ collisions with 10 GeV center-of-mass energy at the Cornell 
Electron Storage Ring (CESR).  We find with $90\%$ confidence the upper limit
on the branching fraction $B(\eta \rightarrow e^{+}e^{-}) < 7.7\times
10^{-5}$.  The application of conventional elementary particle theory
to this decay predicts a branching fraction of about $10^{-9}$.
\end{abstract}
\begin{center}
T.~E.~Browder,$^{1}$ F.~Li,$^{1}$ Y.~Li,$^{1}$
J.~L.~Rodriguez,$^{1}$
T.~Bergfeld,$^{2}$ B.~I.~Eisenstein,$^{2}$ J.~Ernst,$^{2}$
G.~E.~Gladding,$^{2}$ G.~D.~Gollin,$^{2}$ R.~M.~Hans,$^{2}$
% ...
\end{center}

{\em\small 
\begin{center}
$^{1}${University of Hawaii at Manoa, Honolulu, Hawaii 96822}\\
$^{2}${University of Illinois, Champaign-Urbana, Illinois 61801}\\
% ...
$^{14}${Southern Methodist University, Dallas, Texas 75275}\\
% ...
\end{center}}

\newpage

\section{Introduction}

      We have used the CLEO II detector at the Cornell Electron Storage
Ring (CESR) to study about $2\times 10^{7}$ events of the form
$e^{+}e^{-} \rightarrow $ hadrons to search for the rare decay mode
$\eta \rightarrow e^{+} + e^{-}$.
We have not found this decay, only a $90\%$ confidence upper limit of
$B(\eta \rightarrow e^{+}e^{-}) < 7.7\times 10^{-5}$.  As discussed in
Sec. II on the conventional theory of this decay, the predicted branching
fraction is about $10^{-9}$.  An observation of a signal above this level
could be evidence for an unconventional process which enhances the
$\eta \rightarrow e^{+}e^{-}$ decay rate.  

      The plan of this paper is as follows.  Section II gives the
conventional theory for the decay and the predicted relationship between
$B(\eta \rightarrow e^{+}e^{-})$ and the measured branching fraction 
$B(\eta \rightarrow \mu^{+} \mu^{-})$.  Then Sec.\ III describes the data
and the calculation of the number of $\eta$'s produced in the $2\times
10^{7}$ hadron events.  Section IV describes the method 
used to search for $\eta \rightarrow e^{+}e^{-}$ decays.  Finally
Sec.\ V contains the calculation of the upper limit on $B(\eta \rightarrow
e^{+}e^{-})$, a discussion of the errors, and some general remarks on the
search.

\section{Conventional Theory for $\eta \rightarrow \lowercase{e^{+}e^{-}}$ 
and $\eta \rightarrow \mu^{+} \mu^{-}$}

\begin{figure}[htb]
\centerline{
\psfig{figure=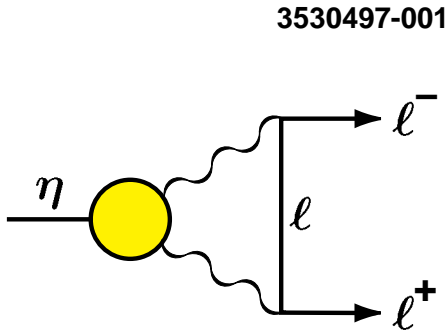,width=2.5in}}
\caption{Feynman diagram for the conventional theory of the decays
$\eta \rightarrow e^{+}e^{-}$ and $\eta \rightarrow \mu^{+} \mu^{-}$.
$l$ represents an $e$ or a $\mu$.}
\label{fig1}
\end{figure}

      The $\eta$ is massive enough (547 MeV/c$^{2}$) to decay via $\eta
\rightarrow \mu^{+} \mu^{-}$ as well as via $\eta \rightarrow e^{+}e^{-}$.
Figure\ \ref{fig1} shows the decay mechanism according to conventional
theory \cite{a,b}.  The decay matrix element for $\eta$ into two
virtual photons, represented by the cross hatched circle, is difficult
to calculate from first principles.  Indeed the same problem occurs in
the study of the decay $\pi^{0} \rightarrow e^{+}e^{-}$ \cite{a,b}.

      Landsberg \cite{b} has reviewed the conventional theory for the decay
$P \rightarrow l^{+}l^{-}$ where $P$ is a pseudoscalar meson, and he
gives the formula for the branching fraction
\begin{equation}
B(P \rightarrow l^{+}l^{-}) = B(P \rightarrow \gamma \gamma)2\alpha^{2}
r^{2}s[|X|^{2}+|Y|^{2}]. \label{eq1}
\end{equation}
Here $B(P \rightarrow \gamma \gamma)$ is the branching fraction for 
$P \rightarrow \gamma \gamma$, $\alpha$ is the fine structure constant,
$r=m_{l}/m_{P}$ where $m_{l}$ and $m_{P}$ are the $l$ and $P$ masses,
and $s=(1-4r^{2})^{\frac{1}{2}}$.
$Y$ is proportional to the imaginary part of the decay amplitude and has
the explicit form
\begin{equation}
|Y|=\frac{1}{s} \ln \left( \frac{1+s}
{2r} \right).
\end{equation}
$X$ is proportional to the real part of the decay amplitude, and is
difficult to calculate with certainty.  Of course even if $X$ is
calculated precisely, Eq.\ (\ref{eq1}) is still not a basic formula
because $B(P \rightarrow \gamma \gamma)$ must be taken from experiment.
Nevertheless Eq.\ (\ref{eq1}) with $X=0$ gives a lower limit on 
$B(P \rightarrow l^{+}l^{-})$, namely
\begin{equation}
B(P \rightarrow l^{+}l^{-}) > B(P \rightarrow \gamma \gamma) \frac{2\alpha
^{2}r^{2}}{s} \left[ \ln \left( \frac{1+s}{2r} \right) \right] ^{2}.
\label{eq3}
\end{equation}

     Using the $\eta \rightarrow \gamma \gamma$ branching fraction from the
Particle Data Group\ \cite{c},
\begin{equation}
B(\eta \rightarrow \gamma \gamma) = 0.393\pm 0.003
\end{equation}
and Eq.\ (\ref{eq3}) we can calculate
\begin{equation}
B(\eta \rightarrow \mu^{+} \mu^{-})_{min} = 4.4\times 10^{-6}. \label{eq5}
\end{equation}
The measured branching fraction\ \cite{c,d,e},
\begin{equation}
B(\eta \rightarrow \mu^{+} \mu^{-})_{measured} = (5.7\pm 0.8)\times
10^{-6} \label{eq6}
\end{equation}
is consistent with this limit.
If we assume that the ratio $|X|^{2}/|Y|^{2}$ is the same for
$\eta \rightarrow e^{+}e^{-}$ and $\eta \rightarrow \mu^{+} \mu^{-}$
then
\begin{equation}
\frac{B(\eta \rightarrow e^{+}e^{-})}{B(\eta \rightarrow \mu^{+}
\mu^{-})} = \left( \frac{r_{e}}{r_{\mu}} \right) ^{2} \left[ \frac
{\ln \left( \frac{1+s_{e}}{2r_{e}} \right)}{\ln \left( \frac{1+s_{\mu}}
{2r_{\mu}} \right)} \right] ^{2} \frac{s_{\mu}}{s_{e}} \label{eq7}
\end{equation}
where the $e$ and $\mu$ subscripts refer to $m_{e}$ and $m_{\mu}$ in
$r=m_{l}/m_{P}$.  The dominant term, $(r_{e}/r_{\mu})
^{2} = (m_{e}/m_{\mu})^{2}$, is due to helicity suppression.  From
Eq.\ (\ref{eq7}), 
\begin{equation}
\frac{B(\eta \rightarrow e^{+}e^{-})}{B(\eta \rightarrow \mu^{+} \mu^{-})}
= 4.05\times 10^{-4}. \label{eq8}
\end{equation}
Eq.\ (\ref{eq8}) also holds for $B(\eta \rightarrow e^{+}e^{-})_{min}/
B(\eta \rightarrow \mu^{+} \mu^{-})_{min}$.  Using this and Eq.\ (\ref
{eq5}), 
\begin{equation}
B(\eta \rightarrow e^{+}e^{-})_{min} = 1.8\times 10^{-9}.
\end{equation}
Finally, combining Eqs.\ (\ref{eq6}) and\ (\ref{eq8}),
\begin{equation}
B(\eta \rightarrow e^{+}e^{-}) \approx 2.3\times 10^{-9}.
\end{equation}

      This estimated branching fraction is based on the assumption that
$|X|^{2}/|Y|^{2}$ is the same for the $e^{+}e^{-}$ and $\mu^{+} \mu^{-}$
decays.  An unknown process present in the
decay $\eta \rightarrow e^{+}e^{-}$, but not in the decay $\eta \rightarrow
\mu^{+} \mu^{-}$, could result in a value of $|X|^{2}/|Y|^{2}$ much larger
in the $e^{+}e^{-}$ mode, and thus a signal larger than the above limit.

\section{Data and Number of $\eta$'s Produced}

\subsection{Detector and Data}

      We used data collected by the CLEO II detector\ \cite{f} at CESR.
The components of the detector which are most
critical to this study are the three concentric cylindrical drift chambers
occupying the space 4 cm to 95 cm radially from the beam axis, comprising a
67-layer charged-particle tracking system which is immersed in a 1.5 Tesla
solenoidal magnetic field.  The momentum $p$, in GeV, of charged particles is
measured with a resolution of $\sigma_{p}/p(\%) \approx [(0.15p)^{2} +
(0.5)^{2}]^{\frac{1}{2}} \approx 0.5\%$ for the electrons we see in
$\eta$ decay.  In addition, ionization loss $(dE/dx)$ is measured
in the 51-layer main drift chamber with a resolution of 6-7$\%$.

      Also important for the measurements reported here is an electromagnetic
calorimeter consisting of 7800 thallium-doped CsI crystals.  These crystals,
each of dimension $\sim$ 5 cm $\times$ 5 cm $\times$ 30 cm, surround the
tracking volume, convering 98$\%$ of the full solid angle.  Forming the barrel
region of the calorimeter, 6144 tapered crystals are arrayed just inside
the magnet coil at a radius of $\sim$ 1 m in a projective cylindrical
geometry, covering 82$\%$ of the solid angle.  The remaining crystals are
rectangular, and are oriented axially in two end caps, overlapping in
solid angle with the ends of the barrel.  The barrel region of the
calorimeter achieves energy and angular resolutions for electromagnetically
showering particles of $\sigma_{E}/E(\%) = 0.35/E^{0.75} + 1.9 - 0.1E$ and
$\sigma_{\phi}($mrad$) = 2.8/\sqrt{E} + 2.5$ ($E$ in GeV), respectively.  The
resulting photon energy and direction information provided by this system
is used to reconstruct $\eta \rightarrow \gamma \gamma$ decays to determine
our $\eta$ sample size as described below.  The crystals are also used in
distinguishing electrons from pions.  When combined with the tracking and
$dE/dx$ information from the drift chambers, the misidentification of
pions is limited to less than one percent.

      We used 3.11
fb$^{-1}$ of data at the $\Upsilon$(4S) resonance, 10.57 GeV, and 1.69
fb$^{-1}$ of data below the $\Upsilon$(4S) resonance at 10.53 GeV.  The 
samples of on resonance and off resonance hadronic events were:
\begin{equation}
N_{had}($on$) = 1.78\times 10^{7}
\end{equation}
\begin{equation}
N_{had}($off$) = 0.797\times 10^{7}.
\end{equation}
    
      We note that 75$\%$ of the on resonance events are of the same type
as the off resonance (also called continuum) events.  The remaining
25$\%$ of the on resonance events are from $e^{+}e^{-} \rightarrow
B\bar{B}$ production.  We found that a major
source of contamination in the search for $\eta \rightarrow e^{+}e^{-}$
decay is one real electron from the semileptonic decay of the $B$ or 
$\bar{B}$ plus one false electron from a pion from the $\bar{B}$ or
$B$ decay chain.  We substantially reduce the number of
$B\bar{B}$ events by applying the following selection criterion to the
on resonance events.  We require $R_{2} \geq 0.3$, where $R_{2} \equiv
H_{2}/H_{0}$ and $H_{0}, H_{2}$ are the zeroth and second Fox-Wolfram
moments, respectively\ \cite{g}.  This selectively removes almost all of
the more spherically-shaped events (see Appendix A for 
clarification).  Thus when discussing our total sample size, we
emphasize the number of continuum events in our data sample:
\begin{equation}
N_{had}($on$) = 1.34\times 10^{7} \label{data1}
\end{equation}
\begin{equation}
N_{had}($off$) = 0.797\times 10^{7}. \label{data2}
\end{equation}
Note that this requirement also removes about fifty percent of the more
collimated continuum events; this effect is included in the efficiencies
we determine below.

      We determined the number of $\eta$ mesons in the events in Eqs.\ 
(\ref{data1}) and\ (\ref{data2}) using the following procedure.  We
selected pairs of photons from the decay
\begin{equation}
\eta \rightarrow \gamma \gamma.
\end{equation}
We then used continuum $e^{+}e^{-}
\rightarrow$ hadron events generated from Monte Carlo\ \cite{h} 
and our simulation of the properties of the CLEO II detector to determine the
efficiency for detecting $\eta \rightarrow \gamma \gamma$.  We also calculated
the average number of $\eta$'s produced per event.  We now give the details
of the procedure.

\subsection{Observed $\eta \rightarrow \gamma \gamma$ Decays}

      In developing the criteria for selecting photon pairs from $\eta
\rightarrow \gamma \gamma$ we kept in mind that we would be looking for
$\eta \rightarrow e^{+}e^{-}$.  As much as possible we chose the same 
selection criteria for $\gamma$ pairs as we would use for the
$e^{+}e^{-}$ pairs.  In this way any uncertainties in the Monte Carlo modeling
of the events will be applied to both decays and thus cancel
out of the analysis.
For example, since a crucial identification signal for 
both $\gamma$'s and $e$'s is an electromagnetic shower in the
calorimeter, and since a minimum shower energy of 0.4 GeV is required for
precise $e-\pi$ separation, we set the minimum shower energy at 0.4 GeV
for $\gamma$'s as well.

      Only $e^{+}e^{-} \rightarrow$ hadron events were used by
requiring the following selection criteria.  First, the events must have
at least five charged tracks each of momentum greater than 225 MeV/c.  Second,
non-annihilation events such as those from beam-gas or beam-wall
interactions were rejected.  Third, for events taken at beam energies
corresponding to the $\Upsilon(4S)$ resonance, $B\bar{B}$
events were removed by the $R_{2}$ cut described previously.  Further 
selection criteria for identifying $\gamma$'s from $\eta \rightarrow \gamma
\gamma$ were:

a.  The electromagnetic shower must occur in the main (barrel) portion of
the electromagnetic calorimeter.  Specifically, $|\cos \theta | < 0.71$ where
$\theta$ is the angle between the $\gamma$ direction and the beam axis.

b.  The angle between the shower and the nearest charged particle track must
be larger than $20^{o}$.

c.  The shower energy must be larger than 0.4 GeV.

d.  The pattern of energy deposition of the shower in the crystals must
be characteristic of a single photon (two or more clusters of deposited
energy indicate randomly overlapping photons or a high-momentum $\pi^{0}$).

e.  The shower must not appear to be a fragment from another shower 
or from a charged pion interaction.

f.  If two $\gamma$'s have an invariant mass within $2.5\sigma$
(12.5 MeV/c$^{2}$)
of the $\pi^{0}$ mass, both $\gamma$'s were discarded.

      Using $\gamma$'s selected with these six criteria we then considered
all combinations of pairs subject to the condition that
\begin{equation}
|${\bf p$_{1}$}$ + ${\bf p$_{2}$}$| > 0.8$ GeV/c$
\end{equation}
where {\bf p$_{1}$} and {\bf p$_{2}$} are the vector momenta of the two
$\gamma$'s.  We set this condition to reduce
the number of random pairings of $\gamma$'s.  

\begin{figure}[htb]
\centerline{
\psfig{figure=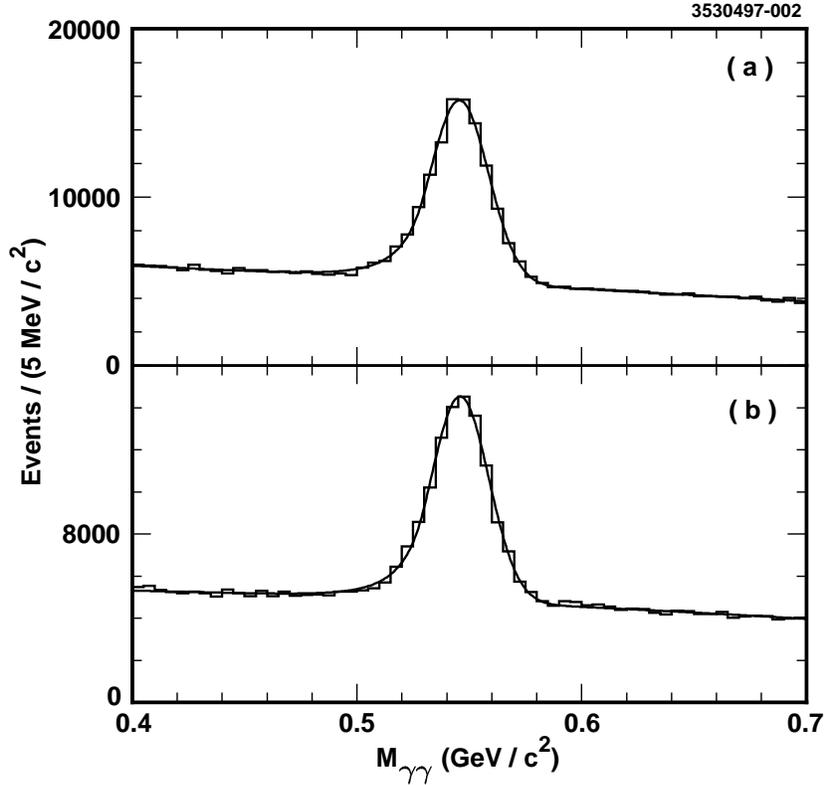,width=5in}}
\caption{Measured $\gamma$ pair invariant mass distribution for (a)
on resonance events and (b) off resonance events.}
\label{fig2}
\end{figure}

      Figure\ \ref{fig2} shows the invariant mass spectra for $\gamma$ pairs
for the on and off resonance data.  The $\eta$ mass peaks at 546
MeV/c$^{2}$ have the widths expected from the properties of the CLEO II
detector, namely $\sigma = 12.8$ MeV/c$^{2}$.  Fits to these spectra gave
the following number of observed $\eta \rightarrow \gamma \gamma$ decays:
\begin{equation}
N_{\eta \rightarrow \gamma \gamma}($on$) = (7.59 \pm 0.12)
\times 10^{4} \label{data3}
\end{equation}
\begin{equation}
N_{\eta \rightarrow \gamma \gamma}($off$) = (6.57 \pm 0.06)
\times 10^{4}. \label{data4}
\end{equation}
Hence we observed about $1.42\times 10^{5}$ $\eta \rightarrow \gamma \gamma$
decays.

\subsection{Efficiency for Detecting $\eta \rightarrow \gamma \gamma$}

      In determining the efficiency for detecting $\eta \rightarrow \gamma
\gamma$ events we restricted ourselved to simulated $e^{+}e^{-}
\rightarrow$ hadron events in the continuum.
Using the selection criteria described in Secs.\ IIIA and IIIB and the 
known number of $\eta$'s in our Monte Carlo sample,
we calculated the efficiency for on resonance (with the $R_{2}$ cut) and
off resonance (without the $R_{2}$ cut).  Since from Eqs.\ (\ref{data1})
and\ (\ref{data2}) we see that $37.5\%$ of our events are off resonance
and $62.5\%$ of our events are on resonance, we use these fractions
to find a weighted mean efficiency for $\eta \rightarrow \gamma \gamma$,
\begin{equation}
\varepsilon_{\eta \rightarrow \gamma \gamma} = (4.80\pm 0.05)\%
\label{gameff}
\end{equation}
where the error is statistical.  The systematic error, which is substantially
larger, will be discussed in Sec. V.  We also found good agreement between the
observed $\eta \rightarrow \gamma \gamma$ events and the simulated 
$\eta \rightarrow \gamma \gamma$ events with respect to the $\eta$
momentum spectrum, the $\eta$ angular distribution, and the $\gamma$ pair
invariant mass distribution of both the $\eta$ peak and the background.

      One major reason for the small efficiency in Eq.\ (\ref{gameff})
is that the $\eta$ momentum, $p_{\eta}$, is required to be larger than 
0.8 GeV/c, but
most $\eta$'s are produced at smaller momenta.  Also responsible for
the small
efficiency is the requirement that the $\gamma$ shower energy in the
calorimeter, $E_{\gamma}$, be greater than 0.4 GeV.  Smaller lower limits
on $p_{\eta}$ and $E_{\gamma}$ would increase $\varepsilon_{\eta \rightarrow
\gamma \gamma}$ substantially, but would result in large increases in
background for $\eta \rightarrow e^{+}e^{-}$.

\section{Search for $\eta \rightarrow \lowercase{e^{+}e^{-}}$ Decays}

      As already noted we limited our systematic uncertainties, particularly
our dependence on simulated event sets, by using as much as possible the same
selection criteria for $\eta \rightarrow e^{+}e^{-}$ events as we used for
the observed $\eta \rightarrow \gamma \gamma$ events, Secs.\ IIIB and IIIC.

      Beginning with the same event sample in Eqs.\ (\ref{data1}) and\
(\ref{data2}) we again require five charged tracks, classification
as an annihilation event, and $R_{2} > 0.3$ for the on resonance events.  We
then looked for showers associated with a charged track which met the
following criteria:

a.  The shower angle $\theta$ must satisfy $|\cos \theta| < 0.71$.

b.  The shower energy must be larger than 0.4 GeV.

c.  The energy deposition of the shower in the crystals must be
characteristic of a single photon.

d.  The shower must not appear to be a fragment from another shower or from
a charged pion interaction.

e.  We require $|${\bf p$_{1}$}$ + ${\bf p$_{2}$}$| > 0.8$ GeV/c.

      The charged particle track had to satisfy our standard criteria for a
track from the primary interaction vertex, namely:

f.  The track had to be of good quality, as identified by the CLEO software
tracking algorithms.

g.  The distance of closest approach of the track to the beam line had to
be less than 5 mm.

h.  The distance of closest approach of the track to the event vertex measured
parallel to the beam line had to be less than 50 mm.  

Next the track had to be identified as an electron using a standard
CLEO algorithm combining $E/p$, shower shape, and several other parameters.
The algorithm has an efficiency of greater than $90\%$, with the exact
efficiency depending on {\bf p$_{e}$}, and a fake rate from
charged pions of about $0.5\%$.

\begin{figure}
\centerline{
\psfig{figure=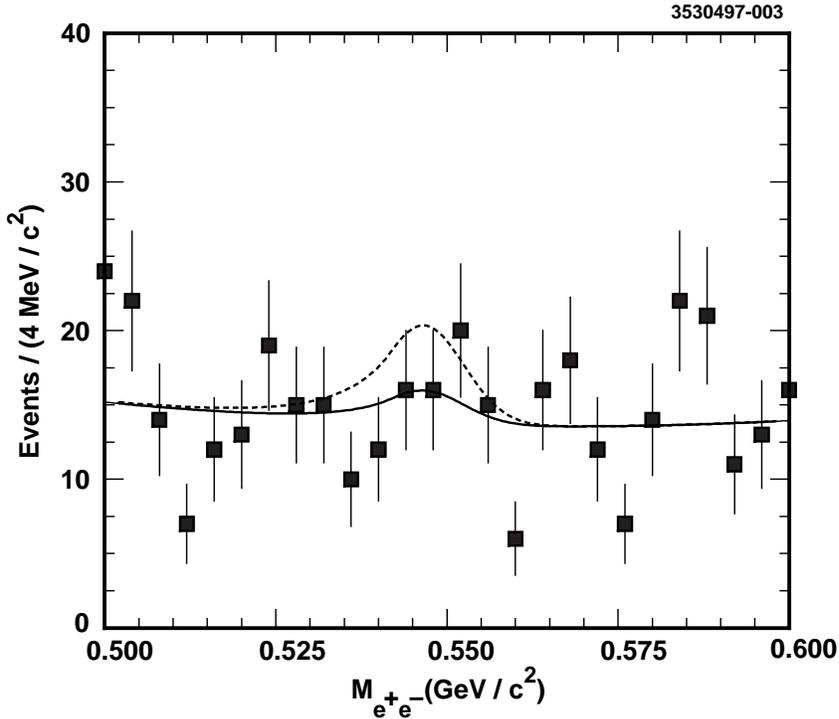,width=5in}}
\caption{Invariant mass spectrum of $e^{+}e^{-}$ pairs found in
the search for $\eta \rightarrow e^{+}e^{-}$ decays.  There is no peak at the
$\eta$ mass of 0.547 GeV/c$^{2}$.  The dashed curve shows the $90\%$ upper
limit for the $\eta \rightarrow e^{+}e^{-}$ signal plus background.}
\label{fig3}
\end{figure}

      We then calculated the invariant mass of every $e^{+}e^{-}$ combination
in the events of Eqs.\ (\ref{data1}) and\ (\ref{data2}).  Figure\
\ref{fig3} shows the spectrum in the mass range of 0.5 to 0.6 GeV/c$^{2}$.
There is no peak at the $\eta$ mass of 547 MeV/c$^{2}$.  A study of
simulated $\eta \rightarrow e^{+}e^{-}$ decays showed that a peak would
have a $\sigma$ of about 5 MeV/c$^{2}$.

\section{Calculation of Upper Limit, Errors, and Final Remarks}

\subsection{Calculation of Upper Limit on $B(\eta \rightarrow e^{+}e^{-})$
and Errors}

      To determine the upper limit on the branching fraction $B(\eta
\rightarrow e^{+}e^{-})$ we have to know the efficiency for the detection
of $\eta \rightarrow e^{+}e^{-}$ decays using the criteria in Sec.\ IV.  
We generated simulated $\eta \rightarrow e^{+}e^{-}$ decays, applied these
criteria and found the total efficiency by taking a weighted mean of on
resonance and off resonance data as in Sec.\ IIIC.  The mean efficiency
was found to be
\begin{equation}
\varepsilon^{\prime}_{\eta \rightarrow e^{+}e^{-}} = (5.22\pm 0.31)\% 
\label{eeff}
\end{equation}
where the error is statistical.

      We used the same simulated events to parameterize the shape of a
hypothetical $\eta \rightarrow e^{+}e^{-}$ peak, resulting in a mass
of $(545.9\pm 0.1)$ MeV/c$^{2}$ and a width $(\sigma)$ of $(5.4\pm 0.1)$ 
MeV/c$^{2}$, where the errors are statistical.  
This shape was then used to fit the data of 
Fig.\ \ref{fig3} to find a $90\%$ confidence upper limit on the number of 
$\eta \rightarrow e^{+}e^{-}$
decays, $N_{\eta \rightarrow e^{+}e^{-}}$.  Varying the mean and $\sigma$
of the fit by one standard deviation, and alternately applying linear
and quadratic background functions, gave a range of values for
$N_{\eta \rightarrow e^{+}e^{-}}$ from 18.4 to 27.1 events.  We used
the most conservative of these fits and concluded  
\begin{equation}
N_{\eta \rightarrow e^{+}e^{-}} < 27.1. \label{maxn}
\end{equation} 

      It is at this point that we must consider the major sources of
systematic error in our analysis.  As we have already minimized the uncertainty
from the electromagnetic calorimeter selection criteria, our remaining 
sources of systematic error
stem from our tracking, particle idenfication, and photon detection 
efficiencies, as well as
the uncertainties in $N_{\eta \rightarrow \gamma \gamma}$, $\varepsilon
_{\eta \rightarrow \gamma \gamma}$, $\varepsilon^{\prime}_{\eta \rightarrow
e^{+}e^{-}}$, and $B(\eta \rightarrow \gamma \gamma)$.  The total 
systematic uncertainty of $10.7\%$ is calculated in Table\ \ref{t2}.
In our final calculation we reduce our mean efficiency by this amount,
yielding:
\begin{equation}
\varepsilon_{\eta \rightarrow e^{+}e^{-}} = 4.66\%.
\label{aeeff}
\end{equation}

      Finally, we normalize to the branching fraction of $\eta \rightarrow
\gamma \gamma$,
\begin{equation}
B(\eta \rightarrow e^{+}e^{-})= \frac{N_{\eta \rightarrow e^{+}e^{-}}}
{N_{\eta \rightarrow \gamma \gamma}}\times \frac{\varepsilon_{\eta
\rightarrow \gamma \gamma}}{\varepsilon_{\eta \rightarrow e^{+}e^{-}}}
\times B(\eta \rightarrow \gamma \gamma)
\end{equation}
and arrive at our $90\%$ confidence upper limit,
\begin{equation}
B(\eta \rightarrow e^{+}e^{-})< 7.7\times 10^{-5}.
\end{equation}
This upper limit is indicated by the dashed curve in Fig.\ \ref{fig3}.

\begin{table}[htb]
\caption{Summary of Systematic Uncertainty}
\label{t2}
\medskip

\begin{tabular}{|l|c|}
Source & Uncertainty \\ \hline
Tracking efficiency & 1$\%$ per $e$ candidate \\ 
Electron ID efficiency & 3$\%$ per electron \\
Photon detection efficiency & 3$\%$ per photon\\ 
$N_{\eta \rightarrow \gamma \gamma}$ (stat.) & 1.4$\%$ \\ 
$\varepsilon_{\eta \rightarrow \gamma \gamma}$ (stat.) & 1.0$\%$ \\
$\varepsilon_{\eta \rightarrow e^{+}e^{-}}$ (stat.) & 6.0$\%$ \\
$B(\eta \rightarrow \gamma \gamma)$ & 0.7$\%$ \\ \hline
Total & 10.7$\%$ \\
\end{tabular}
\end{table}

\newpage

\subsection{Final Remarks}

      Our limit of $B(\eta \rightarrow e^{+}e^{-}) < 7.7\times 10^{-5}$
agrees with and improves upon the upper limit of $2\times 10^{-4}$ found
by White {\it et\ al.}\ \cite{i}; both of the confidence levels are $90\%$.  
White {\it et\ al.} used the reaction $p + d \rightarrow ^{3}$He$ + \eta$ 
to produce 
$\eta$'s combined with a two-arm counter telescope to search for the
$\eta \rightarrow e^{+}e^{-}$ decay.  In the course of concluding our
analysis we have considered if improvements could be made in our method.

      We note from Eqs.\ (\ref{data3}) and\ (\ref{data4}) that
about $1.4\times 10^{5}$ $\eta \rightarrow \gamma \gamma$
events were observed.  Since $\varepsilon_{\eta \rightarrow \gamma 
\gamma}$ and
$\varepsilon_{\eta \rightarrow e^{+}e^{-}}$ are about the same, we
should have been able to investigate a $B(\eta \rightarrow e^{+}e^{-})$
of the order
\begin{equation}
B(\eta \rightarrow e^{+}e^{-}) \sim \left( \frac{2.3}{1.4\times 10^{5}}
\right) B(\eta \rightarrow \gamma \gamma) \sim 6\times 10^{-6}
\end{equation}
if there were no background events.  In the future when the number of
detected $e^{+}e^{-} \rightarrow$ hadron events increase twofold or more
at CESR, and at the B-factories now under construction, one might hope to
achieve a sensitivity of $10^{-6}$.

      However there is a background primarily from pairs containing one
true electron and one pion misidentified as an electron.  Removing $B\bar{B}$
decays aided us somewhat, but unless this background is further reduced,
sensitivities of $10^{-5}$ to $10^{-6}$ for $B(\eta \rightarrow e^{+}e^{-})$
cannot be achieved by our method.  The RICH detector that will be
installed for CLEO III should help improve the pion-electron separation.
In addition, substantial improvement
in sensitivity will probably be achieved using fixed target $\eta$ 
production via
hadronic collisions and specially designed electronic detectors.

\section{Acknowledgements}

	We gratefully acknowledge the effort of the CESR staff in providing 
us with excellent luminosity and running conditions.
J.P.A., J.R.P., and I.P.J.S. thank                                           
the NYI program of the NSF, 
M.S. thanks the PFF program of the NSF,
G.E. thanks the Heisenberg Foundation, 
K.K.G., M.S., H.N.N., T.S., and H.Y. thank the
OJI program of DOE, 
J.R.P., K.H., M.S. and V.S. thank the A.P. Sloan Foundation,
A.W. and R.W. thank the 
Alexander von Humboldt Stiftung,
and M.S. thanks Research Corporation
for support.

\newpage

\appendix

\section{Validity of Supressing $B\bar{B}$}

      Our efficiencies for the on resonance and off resonance searches for
$\eta \rightarrow \gamma \gamma$, i.e. with and without the $R_{2}$ cut,
respectively, are
\begin{equation}
\varepsilon_{\eta \rightarrow \gamma \gamma}($on$) = (3.75
\pm 0.05)\% \label{on}
\end{equation}
\begin{equation}
\varepsilon_{\eta \rightarrow \gamma \gamma}($off$) = (6.57
\pm 0.07)\%. \label{off}
\end{equation}
Combining Eq.\ (\ref{off}) with the number of events in our off
resonance sample [Eq.\ (\ref{data2})], the number of $\eta$'s
observed [Eq.\ (\ref{data4})], and $B(\eta \rightarrow \gamma \gamma)$, 
we find the number of $\eta$'s produced per continuum event to be:
\begin{equation}
n_{\eta}($cont.$) = 0.315\pm 0.006.
\end{equation}
Multiplying this result by the number of on resonance continuum events
[Eq.\ (\ref{data1})], the on resonance efficiency [Eq.\ (\ref{on})],
and $B(\eta \rightarrow \gamma \gamma)$ gives the number of $\eta \rightarrow
\gamma \gamma$ decays we should expect to see from on resonance continuum
events,
\begin{equation}
N_{\eta \rightarrow \gamma \gamma}($on$) = (6.50\pm 0.25)\times
10^{4}.
\end{equation}
Subtracting this from the number of decays we do see, Eq.\ [(\ref{data3})],
this leaves about $10900\pm 2800$ $\eta \rightarrow \gamma \gamma$ decays
that must come from $B\bar{B}$ events, or about $(8\pm 2)\%$ of the
total number of on and off resonance decays we observe.  Thus it seems
that our assumption that all $B\bar{B}$ events are supressed leads
to an small overestimate of $N_{\eta \rightarrow \gamma \gamma}$.

      However, one must remember that $N_{\eta \rightarrow \gamma \gamma}$
and $N_{\eta \rightarrow e^{+}e^{-}}$ will have almost equal proportional
contributions from $B\bar{B}$ events.  In fact the only discrepancy
between the two contributions will be due to slightly different acceptances,
caused by the differing angular and momentum distributions of the two
channels.  Since in calculating our final limit we take the ratio
$N_{\eta \rightarrow e^{+}e^{-}}/N_{\eta \rightarrow \gamma \gamma}$, the
$B\bar{B}$ contributions will almost entirely cancel.  The remaining effect
will be much smaller than $8\%$, and thus negligible for an upper limit.

% now the references. delete or change fake bibitem. delete next three
%   lines and directly read in your .bbl file if you use bibtex.


\begin{references}
\bibitem{a}L. Bergstr\"{o}m, Z. Phys. C {\bf 14}, 129 (1982).
\bibitem{b}L. G. Landsberg, Phys. Reports {\bf 128}, 301 (1985).
\bibitem{c}{\it Review of Particle Properties}, Phys. Rev. {\bf D54},
1 (1996).
\bibitem{d}R. S. Kessler {\it et.\ al.}, Phys. Rev. Lett. {\bf 70}, 892 (1993).
\bibitem{e}R. I. Dzholyadin {\it et\. al.}, Phys. Lett. {\bf 97B}, 471 (1980).
\bibitem{f}CLEO Collaboration, Y. Kubota {\it et.\ al.}, Nucl. Instrum. Methods
Phys. Res, Sect A {\bf 320}, 66 (1992). 
\bibitem{g}G. Fox and S. Wolfram, Phys. Rev. Lett. {\bf 41}, 1581 (1978).
\bibitem{h}JETSET 7.3: T. Sj\"{o}strand and M. Bengtsson, Comput. Phys.
Commun. {\bf 43}, 367 (1987).
\bibitem{i}D. B. White {\it et\ al.}, Phys. Rev. {\bf D53}, 6658 (1996).
\end{references}
\end{document}